\documentclass[twoside,pra,a4paper,10pt,tightenlines,twocolumn,fleqn,showpacs]{revtex4}
\usepackage{amsmath,amssymb}
\newif\ifpdf
\ifx\pdfoutput\undefined
\pdffalse 
\else
\pdfoutput=1 
\pdftrue
\fi
\ifpdf
\usepackage[pdftex]{graphicx}
\else
\usepackage{graphicx}
\fi
\def\vec#1{{\mathbf{#1}}}
\begin{document}
\title{Mass-symmetry breaking in three-body ions}
\author{Vladimir Korobov}
\altaffiliation{Permanent address: Joint Institute for Nuclear Research,
141980, Dubna, Russia}
\author{J.-M.~Richard}
\affiliation{Laboratoire de Physique Subatomique et Cosmologie, Universit\'e Joseph
Fourier -- CNRS-IN2P3\\
53, avenue des Martyrs, F-38026 Grenoble Cedex, France}
\date{\today}
\begin{abstract}
The ground-state energy of  three-body ions $(M^+,M^+,m^-)$ evolves when
the like-charge constituents are given different masses. The comparison of
$(m_1^+,m_2^+,m^-)$ with the average of $(m_1^+,m_1^+,m^-)$ and
$(m_2^+,m_2^+,m^-)$ reveals a competition between the symmetric term and
the antisymmetric one. The former dominates in the Born--Oppenheimer
regime such as the $(p,t,e)$ case, while the latter wins for H$^-$-like
systems with two negative light particles  surrounding a heavy nucleus. A
comparison is also made with the case of baryons in simple quark models
with flavour independence.
\end{abstract}
\pacs{36.10.-k,31.15.Ar.31.15.Md}
\maketitle
\section{Introduction}\label{se:intro}
There are many examples of molecules built out of constituents with
identical charge which might acquire different masses when isotopes become
involved. For instance, the H$_2$ molecule is seen in several variants
such as DT where deuterium and tritium replace  the protons. In the field
of exotic atoms and molecules, many configurations can be envisaged, such
as $(p,\mu^-,\pi^-)$ where two of the constituents have slightly different
masses.

The aim of this paper  is to analyze the  variation of energy  in the
simplest case of the ground state of H$_2^+$-like configurations. There is
already a considerable literature on such systems \cite{Armour93}, and
tables of very accurate variational energies are available. The
mass-dependence of the energy has also been analyzed, and useful
approximate formulas have been proposed. Here, we shall restrict ourselves
to the specific problem  of symmetry breaking.

For consistency, the mass-dependence of  the energy $E(m_1,m_2)$ of the
the $(m_1^+,m_2^+,m^-)$ ion will be analyzed from the published (or
recalculated) values  at several neighbouring  points, or from an estimate
of the first and the second-order perturbation terms calculated with the
wave function of  a central configuration  $(M^+,M^+,m^-)$.

The behaviour of the $(m_1,m_2 ,m)$ systems as a  function of $m_1$ and
$m_2$ for a given third mass $m$ and a given interaction potential was
debated some years ago in the framework of confining quark models with
flavour independence, that is to say, an interaction that does not change
when the mass of any of the quarks is varied, in the same way as the
Coulomb potential governs H$^-$, H$_2^+$, or Ps$^-$.  This note offers the
opportunity to make a comparison between the atomic and hadronic cases.
\section{General considerations}\label{se:gen}
The non-relativistic Hamiltonian is written,  with an obvious notation, as
\begin{equation}\label{eq:Ham}
H=\frac{\vec{p}_1^2}{2m_1}+\frac{\vec{p}_2^2}{2m_2}+\frac{\vec{p}^2}{2m}+V\,,
\end{equation}
where the  potential $V$ is the familiar
$V=r_{12}^{-1}-r_{1}^{-1}-r_{2}^{-1}$.

The energy of any level is an increasing function of each inverse mass
$m_i^{-1}$. This is due to $\vec{p}_i^2$ being positive. For the ground
state, the energy is a concave function of any parameter entering the
Hamiltonian linearly \cite{Thirring79}, in particular an inverse mass. The
result is true for the ground state of any given angular momentum, or for
the sum of the $n$ first levels.

\begin{figure}
\hspace{-6mm}
\includegraphics[width=0.45\textwidth]{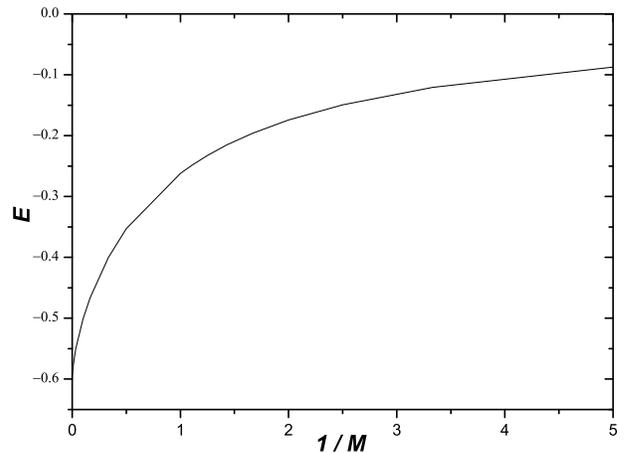}
\caption{\label{FigMMm} Ground-state energy of $(M^+,M^+,m^-)$ as a
function of $m/M$, for $m=1$.}
\end{figure}

In Fig.~\ref{FigMMm}, the energy of the ground state of the symmetric
case, corresponding to  the Hamiltonian
\begin{equation}\label{eq:HamS}
H_\text{S}(M,m)=\frac{1}{2M}[\vec{p}_1^2+\vec{p}_2^2]+\frac{1}{2m}\vec{p}^2+V~,
\end{equation}
is shown as a function of the ratio $x=m/M$, for $m=1$. The monotonic and
concave behaviour is clearly seen. The values of Fig.~\ref{FigMMm} can be
taken from Ref.~\cite{Frolov92}, but we have preferred to recalculate the
energies using a similar variational method employing randomly chosen
complex exponents \cite{Korobov00}. The regular pattern indicates that a
safe interpolation of the binding energy can be done from a few values of
the masses.

The concavity property in Fig.~\ref{FigMMm} shows that, within
\emph{symmetric} configurations, the state $(M^+,M^+,m^-)$ has its energy
above the average of $(m_1^+,m_1^+,m^-)$ and $(m_2^+,m_2^+,m^-)$ if its
inverse mass is taken the exact average, namely,
\begin{equation}
2M^{-1}=m_1^{-1}+m_2^{-1}\,.
\end{equation}

The system of interest is, however, $(m_1^+,m_2^+,m^-)$, and it receives a
downward shift with respect to its symmetrized version $(M^+,M^+,m^-)$.
Indeed, the Hamiltonian (\ref{eq:Ham}) can be rewritten as
\begin{equation}\label{eq:decomp}
H(m_1,m_2,m)= H_\text{S}(M,m)+
\frac{m_1^{-1}\!-\!m_2^{-1}}{2}  (\vec{p}_1^2-\vec{p}_2^2)\>.
\end{equation}
The symmetry-breaking term $\propto  (m_1^{-1}-m_2^{-1})$ lowers the
ground-state energy, as can be seen by applying the variational
principle to (\ref{eq:decomp}), with the symmetric ground state of
$H_\text{S}$ as trial wave function.

In short, when comparing the average of $(m_1^+,m_1^+,m^-)$ and
$(m_2^+,m_2^+,m^-)$ to the mixed state $(m_1^+,m_2^+,m^-)$, there is a
conflict between the concave behaviour in the average inverse mass of the
positive particles, and the downward shift due to symmetry breaking.

Starting from the configuration $(M^+,M^+,m^-)$, one can write the
expansions
\begin{eqnarray}\label{eq:expan}
E(m_i,m_i)&=&E(M,M)+\lambda_i\epsilon_{1,s}+\lambda_i^2 \epsilon_{2,s}+\cdots\\
E(m_1,m_2)&=&E(M,M)+\bar{\lambda}\epsilon_{1,s}+\bar{\lambda}^2\epsilon_{2,s}
+\mu^2\epsilon_{2,a}+\cdots  \nonumber
\end{eqnarray}
where the common third mass is omitted and
\begin{eqnarray}\displaystyle
\lambda_i&=&(2m_i)^{-1}\!-\!(2M)^{-1},\quad
\bar{\lambda}=(\lambda_1+\lambda_2)/2~, \nonumber\\
\mu&=&(4m_1)^{-1}\!-\!(4m_2)^{-1}~,
\end{eqnarray}
which is reduced to $\bar{\lambda}=0$, $\lambda_1=-\lambda_2=\mu$, if
$M^{-1}$ is taken as the exact average of inverse masses $m_1^{-1}$ and
$m_2^{-1}$. In the latter case these expansions reduce to
\begin{equation}\label{eq:expan2}
\begin{array}{@{}l}
E(m_1,m_2)=E(M,M)+\mu^2 \epsilon_{2,a}+\cdots
\\[1.5mm]\displaystyle
\frac{E(m_1,m_1)\!+\!E(m_2,m_2)}{2}=E(M,M)+\mu^2\epsilon_{2,s}+\cdots
\end{array}
\end{equation}

The coefficients $\epsilon$ can be estimated either by fitting the energies
computed for the neighbouring  values of the constituent masses, or  estimated from
the wave function of $(M^+,M^+,m^-)$, using perturbation theory. The
general results listed above translate into $\epsilon_{1,s}>0$ for each
level, and $\epsilon_{2,s} <0$ and $\epsilon_{2,a} <0$ for the ground
state in any sector of conserved quantum numbers.

\section{Results}\label{se:resu}
\subsection{Stability}
It can be shown analytically \cite{Hill77} that $(M^+,M^+,m^-)$ is stable
against dissociation into $(M^+,m^-)$ and an isolated $M^+$, for any value
of the mass ratio $M/m$. The binding energy
\begin{equation}
E_B=E(M^+,M^+,m^-)-E_\text{th}(M^+,m^-)
\end{equation}
improves dramatically as $M$ increases, in the region where $M>m$.
However, $E_B/E_\text{th}$ is not monotonic in the full range of values of
$M/m$.  In Fig.~2, it is seen that the binding energy reaches its minimum
at $M\approx0.5$ and then starts to increase again when the system goes
to the $\mbox{H}^-$ limit.
\begin{figure}\label{fig:Eb}
\hspace{-6mm}
\includegraphics[width=0.45\textwidth]{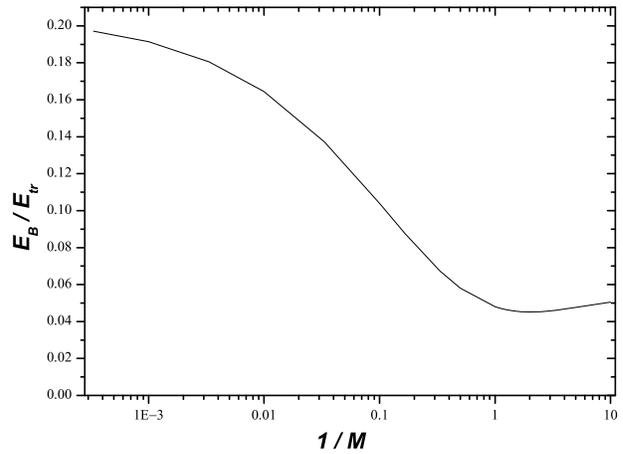}
\caption{Binding energy, $E_B$, of $(M^+,M^+,m^-)$ as a function of
$m/M$ for $m=1$.}
\end{figure}
\subsection{Concavity vs.\ symmetry breaking}
In order to analyze the behaviour of $E(m_1,m_2,m)$ around the central
configuration $E(M,M,m)$ (see Eq.~(\ref{eq:expan})) we introduce the two
operators which determine the coefficients of expansion in
(\ref{eq:expan}), namely,
\begin{equation}\label{eq:Vsa}
V_s = \vec{p}_1^2+\vec{p}_2^2~,\quad\text{and}\quad
V_a = \vec{p}_1^2-\vec{p}_2^2~.
\end{equation}
The perturbation theory gives
\begin{equation}
\begin{array}{@{}l}
\epsilon_{1,s} = \left\langle\Psi_0|V_s|\Psi_0\right\rangle,
\\[3mm]
\epsilon_{2,s} = \left\langle\Psi_0|
                       V_sQ(E_0-H_0)^{-1}QV_s
                    |\Psi_0\right\rangle,
\\[3mm]
\epsilon_{2,a} = \left\langle\Psi_0|
                       V_a(E_0-H_0)^{-1}V_a
                    |\Psi_0\right\rangle,
\end{array}
\end{equation}
where $Q=I-|\Psi_0\rangle\langle\Psi_0|~$ is the projection operator into
the subspace orthogonal to $|\Psi_0\rangle$, and $H_0$, the symmetric
Hamiltonian $H_S(M,m)$ with $m=1$. The first order perturbation for
$V_a$ vanishes.

\begin{table}
\begin{center}
\begin{tabular}{rcccc}
\hline\hline
$M^+$ & $E$ & $\epsilon_{1,s}$ & $\epsilon_{2,s}$ & $\epsilon_{2,a}$ \\
\hline
 0.1 & $-$0.04774837464 & 0.008647 & $-$0.001567 & $-$0.009206 \\
 0.2 & $-$0.08729940294 & 0.028959 & $-$0.009631 & $-$0.059668 \\
 0.3 & $-$0.12069565567 & 0.055480 & $-$0.025623 & $-$0.163192 \\
 0.4 & $-$0.14932813533 & 0.085138 & $-$0.048893 & $-$0.314523 \\
 0.5 & $-$0.17418583295 & 0.116148 & $-$0.078225 & $-$0.502118 \\
 0.6 & $-$0.19599515999 & 0.147453 & $-$0.112387 & $-$0.713757 \\
 0.7 & $-$0.21530357938 & 0.178428 & $-$0.150301 & $-$0.938895 \\
 0.8 & $-$0.23253249352 & 0.208710 & $-$0.191087 & $-$1.169321 \\
 0.9 & $-$0.24801209057 & 0.238095 & $-$0.234046 & $-$1.399056 \\
   1 & $-$0.26200507023 & 0.266477 & $-$0.278636 & $-$1.623973 \\
   2 & $-$0.35268735314 & 0.496963 & $-$0.757118 & $-$3.356412 \\
   3 & $-$0.40026703884 & 0.658243 & $-$1.243673 & $-$4.319080 \\
   6 & $-$0.46608749725 & 0.961169 & $-$2.754451 & $-$5.508756 \\
  10 & $-$0.50180015339 & 1.206036 & $-$5.029963 & $-$6.049197 \\
  30 & $-$0.55024630859 & 1.851059 & $-$20.99919 & $-$6.800187 \\
 100 & $-$0.57644241286 & 2.944942 & $-$117.3044 & $-$7.767019 \\
 300 & $-$0.58833372341 & 4.637278 & $-$595.6470 & $-$9.343615 \\
1000 & $-$0.59509329970 & 7.900453 & $-$3598.177 & $-$12.52000 \\
3000 & $-$0.59836899224 & 13.16189 & $-$18659.60 & $-$17.73284 \\
\hline\hline
\end{tabular}
\end{center}
\caption{The first and second order perturbation for the operators $V_s$
and $V_a$ for the ground state of the system
$\{M^+,M^+,m^-\}$ with $m^-=1$.}
\end{table}

Results of numerical calculations are presented in Table I. From these
data one can get, for instance, an approximate value for the $\mbox{HD}^+$ ion
ground state, using data for $M=3000$ and Eq.~(\ref{eq:expan}),
\begin{equation}\label{eq:pdeapp}
E(m_p,m_d)=-0.597901\mbox{ a.u.}~,
\end{equation}
to be compared with the exact solution
\begin{equation}\label{eq:pdeexa}
E_{\text{HD}^+}=-0.597897968\dots~,
\end{equation}
showing that this rough estimate provides about 6 significant
digits for the energy.

From these data one may define approximate bounds for the stability
region by solving the equation,
\begin{equation}\label{eq:stability}
E(M,M)+\mu^2\epsilon_{2,a}=
   -\frac{1}{2}\left(\frac{1}{M}+2\mu+1\right)^{-1},
\end{equation}
with respect to $\mu$. Here we assume that $m_1>m_2$ and $m=1$. The
variation of $\mu$ corresponds to moving along an horizontal line in the
triangle of normalized inverse masses \cite{Marten92}. The value of $\mu$ obtained from
Eq.~(\ref{eq:stability}) for $M=0.1$ is $\mu\approx -0.55$, that
corresponds to $m_1=0.112$ and $m_2=0.0901$, the ratio of these two masses
is $m_1/m_2=1.25$.

\begin{figure}\label{fig:Vs_Va}
\hspace{-6mm}
\includegraphics[width=0.45\textwidth]{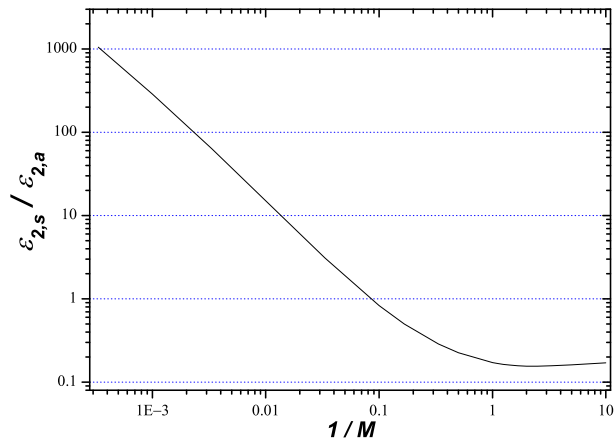}
\caption{The ratio of the second order terms (symmetry vs.\ asymmetry).}
\end{figure}

The ratio of the symmetric to asymmetric second-order terms (i.e., concavity vs.\
symmetry breaking) is shown in Fig.~3.
In the adiabatic region $M\gg m$, the symmetric term dominates. An inverse hierarchy is observed
in the atomic regime $M\ll m$, with the asymmetric term dominating by about
one order of magnitude, implying an inequality
\begin{equation}
2 E(m_1,m_2)<E(m_1,m_1)+E(m_2,m_2)~.
\end{equation}
\subsection{$\mbox{H}_2^+$ ``gerade" and ``ungerade" ground states}
To get a different insight to the problem, we now compare the concavity
and symmetry breaking terms for the $\mbox{H}_2^+$ molecular ion ground
states of different symmetry, starting from the molecule with identical
nuclei. The $2p\sigma_u$ (or ``ungerade") state is bound by a very shallow
potential and lies very close to the threshold. On contrary, the
$1s\sigma_g$ potential well supports the existence of 19 vibrational
states. Hence it is anticipated that opening  the coupling between the
``gerade" and ``ungerade" states lowers the ground $2p\sigma_u$ state more
rapidly than the ground $1s\sigma_g$ state.

Numerical results are presented in Tables II and III. The ``ungerade"
$\epsilon_{2,a}$ exhibits much slower convergence, since the intermediate
state, or the first-order perturbation wave-function, a symmetric function
being  solution for the $1s\sigma_g$ potential at about zero energy, has
multiple nodes that  the variational approximation should reproduce with a
reasonable accuracy. This makes the numerical calculations more delicate.

\begin{table}
\begin{center}
\begin{tabular}{ccc@{\hspace{5mm}}cc}
\hline\hline
$N$ & $\epsilon_{1,s}$ & $\epsilon_{2,s}$ & $N$ & $\epsilon_{2,a}$ \\
\hline
600 & 10.453482 & $-$8940.3037 & 600 & $-$15.043224 \\
800 & 10.453482 & $-$8940.3038 & 800 & $-$15.043224 \\
\hline\hline
\end{tabular}
\end{center}
\caption{The first and second order perturbation for the operators $V_s$
and $V_a$ for the ground $1s\sigma_g$ state of the $\mbox{H}_2^+$.}
\end{table}

\begin{table}
\begin{center}
\begin{tabular}{ccc@{\hspace{5mm}}cc}
\hline\hline
$N$ & $\epsilon_{1,s}$ & $\epsilon_{2,s}$ & $N$ & $\epsilon_{2,a}$ \\
\hline
 600 & 1.0497471 & $-$99.715284 & 1000 & $-$40051.369 \\
 800 & 1.0497471 & $-$99.725303 & 1400 & $-$40342.387 \\
1000 & 1.0497471 & $-$99.725735 & 1800 & $-$40376.809 \\
\hline\hline
\end{tabular}
\end{center}
\caption{The first and second order perturbation for the operators $V_s$
and $V_a$ for the ground $2p\sigma_u$ state of the $\mbox{H}_2^+$.}
\end{table}

The results in the Tables II and III illustrate our conjecture that that
the symmetry breaking effect for the case of $2p\sigma_u$ state is much
stronger than for the ``gerade" ground state, while the concavity term is
almost negligible.
\section{Comparison with baryons}\label{se:baryons}
In Quantum Chromodynamics (QCD), quarks are coupled to gluons through
their colour. The static potential, which serves as first approximation to
describe the interaction,  is thus independent of the mass of the quarks
experiencing this potential. This property, called \emph{flavour
independence}, stimulated several studies on how the spectrum evolves in a
given confining potential, as a function of the quark masses. For
references, see Refs.~\cite{GMbook,Nussinov02}.

%
In the case of mesons, once the centre-of-mass motion is removed, the
interaction Hamiltonian depends only upon the reduced mass $\mu$, and the
inequality \cite{GMbook,Nussinov02}
\begin{equation}\label{eq:Nussinov1}
2E_2(m_1,m_2)\ge E_2(m_1,m_1)+E_2(m_2,m_2)~,
\end{equation}
holds whatever flavour-independent potential $V(r_{12})$ is assumed.

In the case of baryons, it was always found for plausible
flavour-independent potentials \cite{RTAnnals}, that
\begin{equation}\label{eq:Nussinov2}
2E_3(m_1,m_2,m)\ge E_3(m_1,m_1,m)+E_3(m_2,m_2,m)~.
\end{equation}
The approximate equality observed for experimental baryon masses (the
contribution of constituent masses  is identical on both sides) is due
to a cancelation of this concave behaviour with the convex behaviour of
hyperfine corrections.

It was even conjectured that Eq.~(\ref{eq:Nussinov2}) might be generally
true, though resisting  proof. Then Lieb found \cite{Lieb85}
counterexamples, which were further extended in
\cite{Martin86,Nussinov02}. Lieb also found wide classes of potentials for
which the inequality (\ref{eq:Nussinov2}) is rigorously proven. Note that
the quark problem does not include the case of H$_2^+$-like ions, since
the interquark potential $V=\sum v(r_{ij})$ is assumed to be fully
symmetric.

\section{Conclusion}\label{se:conc}
In this article, we have studied the effect of breaking the $m_1=m_2$
symmetry in the ground state $(m_1^+,m_2^+,m^-)$. Two regimes are
identified. For $m_1$ and $m_2$, much heavier than the negatively charged
particle of mass $m$, the heavy particles experience an effective two-body
potential. Then the dynamics depends mainly on the sum of inverse masses,
a property that becomes exact in the case of two-body systems, or in the
case of mesons in a flavour-independent quark--antiquark potential. In
this situation, the inequality (\ref{eq:Nussinov2}) is observed.  Now, for
$m_1$ and $m_2$ of the order of $m$ or smaller, in the $\mbox{H}^-$ limit,
the asymmetric term $(\vec{p}_1^2\!-\!\vec{p}_2^2)$ plays a more important
role, and the reversed inequality is observed.

For the adiabatic systems, which have both symmetric and antisymmetric
bound states, the symmetry breaking effect is much stronger for the upper
antisymmetric state, and is the dominant contribution, while the concavity
term $(\vec{p}_1^2\!+\!\vec{p}_2^2)$ is rather negligible.

We intend to study the case of excited states, and more complicated
structures, such as hydrogen-molecule-like states $(m_1^+,m_2^+,m^-,m^-)$,
and to extend the results of our present work to the higher order
perturbation terms.

\acknowledgments

We thank P.~Valiron for stimulating discussion and M.~Asghar for comments
on the manuscript.


\begin{thebibliography}{100}
\bibitem{Armour93}
E.A.G.~Armour and W.~ Byers Brown, Accounts of Chemical Research, {\bf 26}
168 (1993), and references therein.

\bibitem{Thirring79}
W.~Thirring, {\it A Course in Mathematical Physics}, {\bf Vol.~3}: Quantum
   Mechanics of Atoms and Molecules (Springer Verlag, New-York, 1979).

\bibitem{Frolov92}
A.M.~Frolov and D.M.~Bishop, Phys.\ Rev.~A \textbf{45}, 6236 (1992).

\bibitem{Korobov00}
V.I.~Korobov, Phys.\ Rev.~A {\bf 61}, 064503 (2000).

\bibitem{Hill77}
R.N.~Hill, J.~Math.\ Phys.\ \textbf{18}, 2316 (1977).

\bibitem{Marten92} A.~Martin, J.-M.~Richard, and Tai Tsun Wu, Phys.\ %
Rev.~A \textbf{46}, 3697 (1992).



\bibitem{GMbook}
H.~Grosse and A.~Martin,
{\it Particle physics and the Schr{\"o}dinger equation},
   Cambridge, UK: Univ. Pr. (1997) 167 p. (Cambridge monographs
on particle physics, nuclear physics and cosmology, vol. \#6).

\bibitem{Nussinov02}
S.~Nussinov and M.~A.~Lampert,
Phys.\ Rept.\  {\bf 362} (2002) 193
[arXiv:hep-ph/9911532].

\bibitem{RTAnnals}
J.-M.~Richard and P.~Taxil,
Ann.~Phys.~(N.Y.)\ {\bf 150} (1983) 267.

\bibitem{Lieb85}
E.~H.~Lieb,
Phys.\ Rev.\ Lett.\  {\bf 54}, 1987 (1985).

\bibitem{Martin86}
A.~Martin, J.~M.~Richard and P.~Taxil,
Phys.\ Lett.\ B {\bf 176}, 224 (1986).

\end{thebibliography}
\end{document}